\documentstyle[12pt,epsfig]{article}
\begin{document}
\baselineskip=1.5pc
\def\bfg #1{{\mbox{\boldmath $#1$}}}

\begin{center}
{\large \bf Deuteron Breakup $pd\to pnp$ in Specific Kinematics
and  Short-Range NN-Interaction}
\vskip 0.3cm

Yu.N. Uzikov
\footnote{E-mail address: uzikov@nusun.jinr.dubna.su}\\
{\it  Joint Institute for Nuclear Research, LNP, Dubna, Moscow region 141980,
 Russia}\\
\end{center}
\vskip 0.3cm
 {\bf Abstract.}
   Deuteron breakup
 processes $p+d\to (NN)_{s,t}+N$
 are studied in the kinematics of backward elastic pd-scattering
 at law relative momenta of the NN-pair
 $k\sim  0-50 MeV/c$ for initial energies $T_0=0.5-2.5 GeV$
 in the framework of the known mechanisms  of the $pd\to dp$ process
 including rescatterings in the initial and final states.
  A considerable suppression of the  $\Delta-$ and $N^*-$ isobar
 excitation  mechanisms is found for the production of the
 singlet $(NN)_s$-pair. As a result, one nucleon exchange mechanism
 can be identified in  the cross section and polarization observables
 by the node of  the half-off-shell  NN$(~^1S_0)$-scattering amplitude,
 $t(q,k)$, at $q\sim 0.4 $ GeV/c.



\eject

The short-range structure of the lightest nuclei
 is related to  fundamental problems of the theory of strong interactions.
  At present the main task  consists  in a clear experimental
 observation of this structure  and determination to what  limiting values
 of  the intrinsic nucleon momenta  $q$  the traditional description
 of nuclei in terms of nucleons is valid.
In the framework of  the impulse approximation (IA),
  available inclusive experimental data on spin averaged cross section
 for   deuteron disintegration
  ${ d}\,p\to p(0^o)\,X$, ${ d}\,^{12}C\to p(0^o)\,X$ (see Refs.
\cite{dubnainclus,dubna} and references therein) are
 compatible with  the realistic deuteron wave functions  at  low
 internal momenta  $q<0.3 GeV/c$.
  However, at higher momenta  a systematic deviation from the IA is
 observed \cite{pedrisat90}.  This disagreement is   most obvious  in the
  tensor analyzing power $T_{20}$ of these reactions.
 One possible explanation of such a behaviour is the
 presence of quark exchanges effects
 in the deuteron wave function \cite{kob98}.
  However, other  interpretations have been proposed also \cite{lykasd}.
 The experimental data on exclusive
  polarized and unpolarized experiments  $p+d\to p+n+p$
 demonstrate  similar deviation from the  IA
 which  can be qualitatively explained by  rescatterings in the final state and
  excitation of the $\Delta-$isobar \cite{aleshin}.
 A large contribution from final state interaction  and excitation of
 the nucleon isobars  can produce fast
 nucleon ``spectator'' in final state
 without the necessity of high momentum  pn- components
 of the deuteron wave function and therefore mask the  structure of the
 deuteron at short NN- distances.

 At present a correct treatment of these
 screening effects at intermediate energies
 is a nontrivial problem for  theory.
 This can be seen, for example,
 from the analysis of the backward elastic $pd$-scattering.
 Exact Faddeev calculations  are
 performed at present for the elastic Nd-scattering at energies
 below the pion production threshold  ($T_N\le 250$ MeV)
 \cite{gloeckle}. At higher energies
 analysis was performed in the
 framework of the coherent sum of the one nucleon exchange (ONE),
 $pN$ single scattering (SS) and
 the $\Delta-$ isobar excitation ($\Delta $) mechanisms
\cite{lak,imuzsh89}, \cite{uz98}.
There is an agreement with the experimental data both for $d\sigma/d\Omega$
 and  $T_{20}$   at rather low energies $T_0 \sim 0.2- 0.4GeV$ where
 the ONE-mechanism dominates \cite{imuzsh89}.
 At initial energies $T_0=0.5-1.5 GeV$ and scattering angle
 $\theta_{c.m.}=180^o$  the ONE+ SS+$\Delta $ model qualitatively
  describes  the spin-averaged $pd\to dp$  cross section $d\sigma/d \Omega$
\cite{uz98} but does not explain the $T_{20}$-data
\cite{arvieux, azhg97}.True negative sign of $T_{20}$  was obtained in
 Ref. \cite{bdillig} under special approximations
  in the framework of $\Delta$-mechanism  but the detailed
 structure of $T_{20}$ was not explained.     Inclusion of
  rescatterings
 and  $NN^*$ components of the deuteron wave function \cite{uz97}
 or use of the covariant relativistic approach within the ONE
 and pion exchange  mechanisms \cite{kaptaripdpd}
 does not improve the agreement with the data at $T_0>0.5 GeV$.
\begin{figure}[t]
\mbox{\epsfig{figure=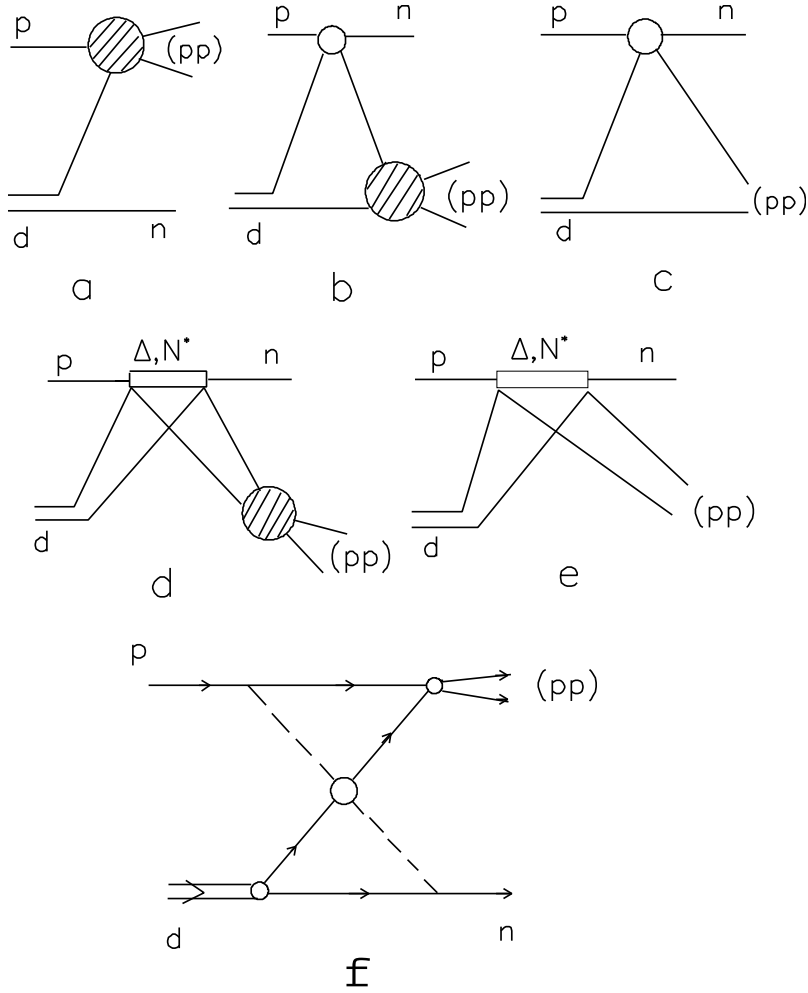,height=0.6\textheight, clip=}}
\caption
{
  Mechanisms of the reaction $p+d\protect\to n+pp$:
       $ a $  -- one-nucleon exchange (ONE),
       $ b, c$ -- single scattering (SS),
       $ d, e$ -- double pN-scattering with excitation
of the $\Delta-$  $(\Delta )$ or $N^*$-isobar.
}
\label{uzi-1}
\end{figure}

 Obviously, this problem is connected to the nucleon isobar
 excitation. It could  be related not only to  off-shell effects
 in the $NN\rightleftharpoons N\Delta$ amplitudes. In fact,
 even in the $\Delta-$region spin observables for the elementary reaction
 $pp\to pn\pi^+$  are not described properly
 within the one-meson-exchange model, as was shown
 in the unitary theory of  coupled $NN-\pi NN$ channels \cite{tlee}.

 To minimize  screening effects we propose  to study the deuteron breakup
 process $p+d\to (np)_{s}+p$ with formation of the singlet deuteron
 $(pn)_s$ in kinematics of the backward (quasi)elastic pd-scattering
 or the similar reaction
\begin{equation}
\label{reaction1}
 p+d\to (pp)(0^o)+n(180^o)
\end{equation}
at small relative energy
  of the forward pp-pair ($E_{pp}=0-3 MeV$) \cite{smuz98}.
 The $ONE+SS+\Delta$ model of the
 backward elastic pd-scattering
 developed in Ref. \cite{lak} and
 improved  later in Ref. \cite{imuzsh89} in respect to the
 $\Delta-$ contribution was  used in  Refs.\cite{imuz90}
 to analyze the reaction (\ref{reaction1}) (see Fig.\ref{uzi-1}).
 According to  \cite {imuz90}, the presence of the final NN-pair in the
 $^1S_0$-state   considerably changes
 the relative contribution of the different mechanisms in
 comparison with the  elastic backward pd-scattering. In particularly,
  a suppression of the $\Delta $-mechanism contribution was found in Ref.
  \cite {imuz90}, but not explained.

 We present here  new results of a theoretical study
 of the reaction (\ref{reaction1}).
 Firstly, we show  that the isotopic spin
 suppression factor $\frac{1}{3}$ arising for the
 $\Delta$ mechanism is effective for  broader class of important
 diagrams
 than  established in Refs.\cite{imuz90},\cite{imuz87}. In particular,
 the same factor occurs for the $N^*$-excitation mechanism too.
 Secondly, we analyze   spatial parts of the amplitudes.
 Thirdly, rescattering effects are estimated here in the eikonal
 approximation.
 In addition  spin observables of the first and the second order
 have been calculated here also.

 We will discuss here  the  ratio of the break-up amplitudes
 with formation of the singlet $(np)_s$- and triplet $(pn)_t$-pairs for
 different mechanisms
\begin{equation}
\label{ratio}
 R\equiv
 \frac{|A(p\,d\to (np)_s\,p)|}{|A(p\,d\to (np)_t\, p)|}
=\frac{|A(p\,d\to (np)_s\,p)|}{|f(k)A(p\,d\to d\,p)|},
\end{equation}
where the following relation  from Ref. \cite{bfw}
\begin{equation}
\label{contin}
A(p\,d\to (pn)_t\,p)= f(k)A(pd\to dp),
\end{equation}
is used. The function $f(k)$ in the last equation
 is defined by Eq.(3) in Ref. \cite{bfw}, where
$k$ is  the relative momentum of the final pn-pair.

 Let us discuss  the isotopic spin factors
for the  mechanism
 with intermediate meson-nucleon rescattering \cite{anjos} depicted
 in Fig.\ref{uzi-1}({\it f}). This diagram includes that part of the
 $\Delta-$ (and $N^*$-) mechanism  depicted in Fig.\ref{uzi-1}({\it d,e})), which
 can be described in  one-meson exchange model
 \cite{lak,bdillig, imuzsh89}. Besides, this diagram
  takes into account   a meson-nucleon continuum too.
 For the isovector mesons in Fig.\ref{uzi-1}({\it f})
 the   $pd\to p(np)_{s,t}$ amplitude
 contains  two terms corresponding to   different values
 of the total isotopic spin  of the intermediate meson-nucleon  system,
 $T=\frac{1}{2}$ and $\frac{3}{2}$:
\begin{eqnarray}
\label{isot3}
A(p\,d\to (pn)_s\,p) = { A}^{s}_{T=1/2}+ {A}^{s}_{T=3/2}, \\ \nonumber
A(p\,d\to (pn)_t\,p) = { A}^{t}_{T=1/2}+ { A}^t_{T=3/2}.
\end{eqnarray}
 For  definite values of the total isotopic spin  $T$ of the intermediate
 meson-nucleon system and the isotopic spin of the forward pn-pair, $T_{pn}$,
 one has got the following isotopic spin  structure of the breakup amplitude
$$A_{T}^{s,t}(p\,d\to (pn)_{T_{np}}\,p)=(2T+1)\sqrt{2(2T_{np}+1}
(T_{np}0\frac{1}{2} \frac{1}{2}|\frac{1}{2} \frac{1}{2})\times$$
\begin{equation}
\label{isot3a}
\times\left \{ \begin{array} {ccc} 1&\frac{1}{2}&\frac{1}{2}\\
 1&\frac{1}{2}& T
\end{array}\right \}
\left \{ \begin{array} {ccc} \frac{1}{2}&\frac{1}{2}&T_{np}\\
 \frac{1}{2}&\frac{1}{2}&1
\end{array}\right \} B_T^{s,t}(T_{np}).
\end{equation}
 The Clebsh-Gordan coefficients and 6j-symbols are used here in
standard notations.
The dynamical factor $B_T$ in Eq. (\ref{isot3a}) does not depend
on the z-projections of the isotopic spins.
Assuming the spatial parts of the
wave functions of the singlet and triplet
$pn$ pairs to be the same,
i.e. $B_T^s(T_{np}=1)/B_T^t(T_{np}=0)=1$
 in Eq. (\ref{isot3a}),
  one can find
from (\ref{isot3a}) the following ratios
\begin{eqnarray}
\label{isot4}
\frac{{ A}^{s}_{T=1/2}}{{ A}^{t}_{T=1/2}}=
\frac {{ A}^{s}_{T=3/2}}{{ A}^{t}_{T=3/2}}= -\frac{1}{3}.
\end{eqnarray}
 Using Eq.(\ref{isot4}) one obtains from Eqs.(\ref{ratio}) and (\ref{isot3})
 \begin{equation}
\label{isot6}
R_I^{iv}=\frac{A(p\,d\to (pn)_s\, p)}{A(p\,d\to (pn)_t\, p)} =- \frac{1}{3}.
  \end{equation}
 We stress that owing to Eq. (\ref{isot4})
 this ratio does not depend on the unknown
  relative phase of the amplitudes
 $A_{1/2}$ and  $A_{3/2}$ of the process $\pi N\to \pi N$.
 The ratio (\ref{isot6}) is  more general  in comparison with the one
   obtained previously in Ref.\cite{imuz87} for the
 $\Delta $- excitation  and one-pion exchange mechanisms \cite{kolybas}.
 In fact, (i)   all intermediate states
 of the  meson-nucleon  system both  for the isotopic spin
 T=3/2 and  T=1/2 are taken into account in Eq. (\ref{isot6})
 including the  $N^*$-pole terms and continuum.
 (ii) Moreover,
  relation (\ref{isot6}) is valid
 not only for one definite isovector meson in the intermediate state
 but for the sum of  diagram in Fig.\ref{uzi-1}({\it f})
 with  the different isovector mesons ($\pi,\, \rho , \dots  $) in
 any combinations ($\pi \pi, \pi\rho, \rho \rho , \dots $).

 In fact, the relative suppression of  the time inverted process
 $p+(NN)\to d+N$ on the singlet NN-pair
 was observed  in the reactions of quasielastic knockout of
 fast deuterons  $^{6,7}Li (p,Nd)$ \cite{albrecht}
 at initial energy $T_0=670$ MeV.
According to the experimental data \cite{albrecht}, the (p,nd) cross
section for the transitions to the states with destroyed $\alpha$-core in the
$^{6,7}Li$ nuclei is one  order of magnitude smaller in comparison with the
  (p,pd) cross section. This fact was  explained in
 Ref.\cite{imuz87} in the framework of the $\Delta $-resonance mechanism
 as a result of the  suppression of the cross section of the process
 $p+(nn)_s\to d+n$ by the isotopic spin factor 1/9 in comparison with the
process $p+(pn)_t\to d+p$.

 For the isoscalar mesons
($\omega, \ \eta,\ \eta ',\, \dots $) in the intermediate state
in Fig.\ref{uzi-1}({\it f}) and for the ONE mechanism  we find the ratios
$R_I^{is}=R_I^{ONE}=1$.
 For the SS-mechanism
 it is impossible to extract   the
 isotopic factor from Eq.(\ref{ratio}).

 Let us take into account  that
 the spatial parts of the scattering wave functions
 in the triplet $^3S_1$  and
 singlet $^1S_0$ states  are different.
 For the  ONE  mechanism  one can find the following
 ratio for the  spatial parts of  the transition amplitudes into the
  S-states of the $(pn)_{s,t}$-pairs
 \begin{eqnarray}
 \label{xratioone}
R_{X} (ONE)= \frac{A(pd\to
 p+(pn)_s)}{A(pd\to p+(pn)_t)}
=\frac {\sqrt{4\pi m}\,t_s(q',k)}
{(\varepsilon+{\bf q}^2/m)\, u({ q})}\frac{1}{f(k)} \equiv
\frac{t_s(q',k)}{t_t(q,k)},
\end{eqnarray}
where
$u(q)$ is
 the deuteron S-wave function normalized as
$ \int _0^\infty[u^2(q)+w^2(q)]\frac{q^2dq}{(2\pi )^3}=1$,
$\varepsilon $ is the deuteron binding energy; the value
$t_t(q,k)$ is defined by Eq.(\ref{xratioone});
\begin{equation}
\label{half}
t_s(q',k)={-4\pi \int_0^\infty \frac{F_0(q'r)}{q'r}V(r)\psi_s(r)r^2dr}
\end{equation}
 is the
half-off-shell  matrix of $pn$-scattering in the $^1S_0$-state,
  $F_0(z)$ is the Coulomb function regular at the origin, $\psi_s$
is  the $^1S_0$- scattering wave function obtained from the solution of the
 Schr\"odinger equation
with $NN(^1S_0$) -interaction potential
$V(r)$ (plus Coulomb term for the pp-system). This function  is normalized
at $r\to \infty $ as
$\psi_s(r)\to {sin(kr+\delta)}/{kr}$,
where $\delta $ is the  phase shift in the $^1S_0$-state of the pn-system.

For the $\Delta$-mechanism  the following ratio
 of the spatial matrix elements is obtained from Eq. (\ref{ratio})
on the base of Eq. (A.5) from Ref.\cite{imuz90}
\begin{equation}
\label{xratiod}
R_{X}(\Delta )=\frac{\sqrt{4\pi m}\int_0^\infty dr
 \exp{(i{\cal A}r)}u(r)\psi_s(r)}
{\int_0^\infty dr \exp{(i{\cal A}r)}u(r)u(r)}\frac{1}{f(k)}
\equiv \frac{I_s(\Delta)}{I_t(\Delta)}.
\end{equation}
Here $u(r)$ (or $w(r)$) is the deuteron S (D)-wave function
 in configuration space  normalized as
$ \int _0^\infty [u^2(r)+w^2(r)]r^2dr=1$.
 The  variable  ${\cal A}$ in Eq.(\ref{xratiod}) is expressed
in Refs. \cite{lak,imuz90} through  kinematical variables
of the process (\ref{reaction1}), masses of the
nucleon $m$ and the deuteron $m_d$,
 the mass $M_\Delta $ and width $\Gamma_\Delta$ of the $\Delta$-
 isobar, respectively.
 The  ratio  for the D-component of the initial
 deuteron, $w$, can be obtained from Eq.(\ref{xratiod}) by the following
  substitution $u(r)\to w(r)$.
\eject
\begin{figure}[htb]
\mbox{\epsfig{figure=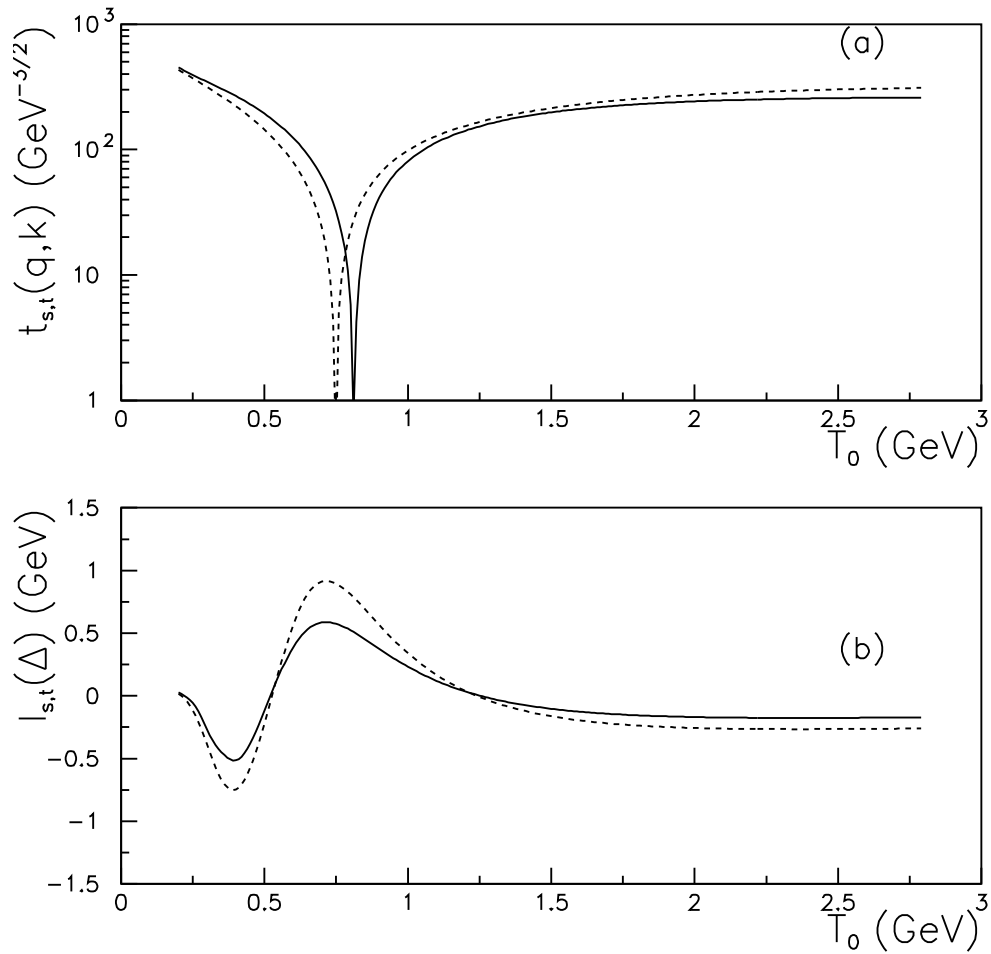,height=0.65\textheight, clip=}}
\caption
{
The numerators (dashed) and denominators (full) in the ratios
 (\protect\ref{xratioone}) for the ONE ({\it a}) and
(\protect\ref{xratiod})
  for the $\Delta- $  mechanism ({\it b}) at $E_{pp}=3$ MeV.
}
\label{xratio}
\end{figure}

 For the $N^*$-mechanism the corresponding
formulas  coincide with Eq.(\ref{xratiod}) except the value
 ${\cal A}$ depends
on the mass $M_{N^*}$ and width $\Gamma_{N^*}$ of the $N^*$-isobar.
 The ratio for
the SS- mechanism can be written in a similar form.

 For the dominating ONE-mechanism
 rescatterings in the initial and final states are taken into account
in the line of works \cite{blu,uz97}. This is an eikonal version of
 the distorted wave Born approximation (DWBA)  method  successfully
 applied in Ref. \cite{blu}  to the backward
 elastic $p^3He$-scattering at $T_0\geq 1 GeV$ in the framework
 of np-pair exchange.  Since the SS- and $\Delta$-mechanisms
 are less important they are considered here in the plane wave
 approximation.
\begin{figure}
\mbox{\epsfig{figure=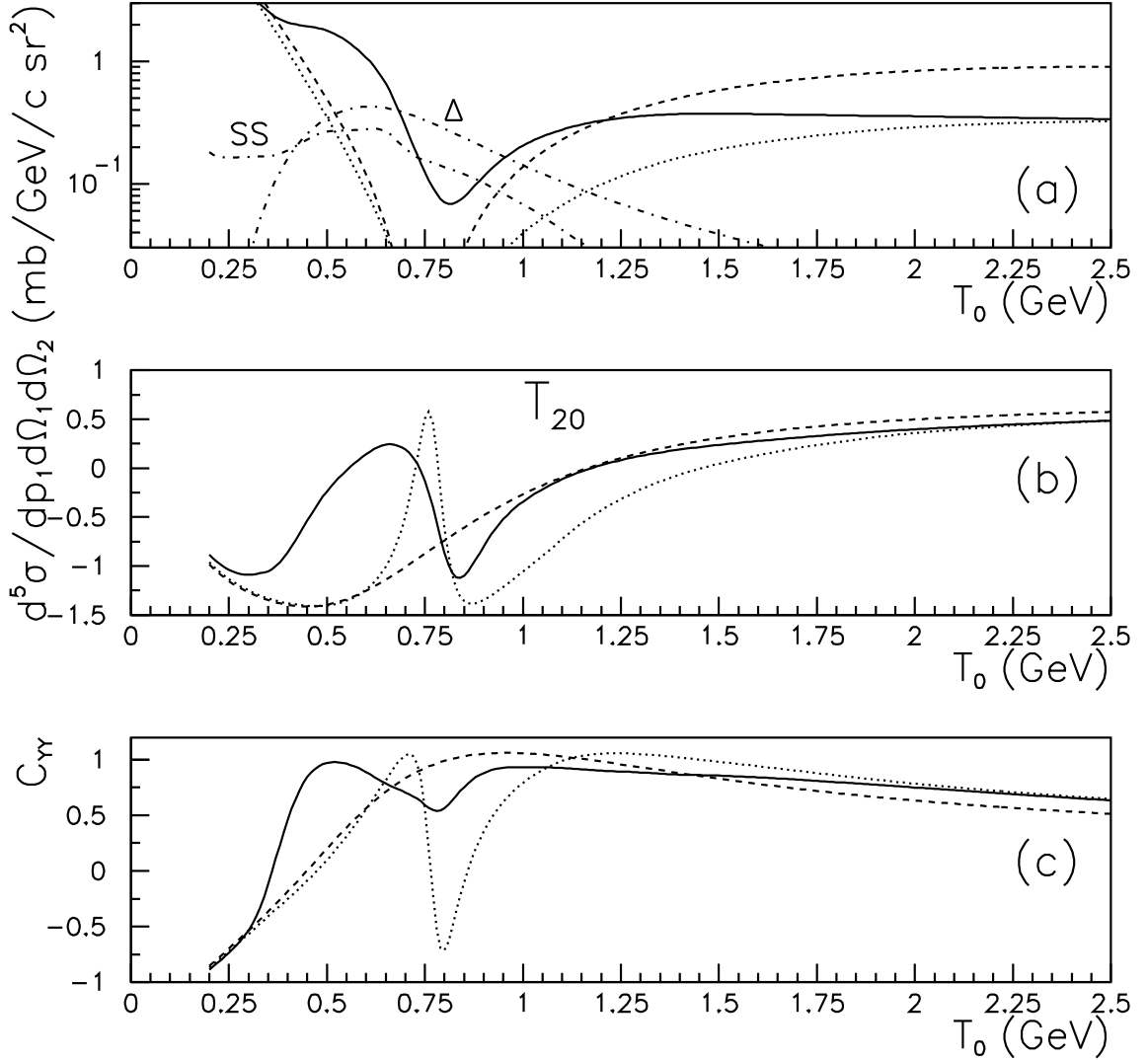,height=0.75\textheight, clip=}}
\caption{
 The lab. system cross section  ({\it a}),
 tenzor analyzing power $T_{20}$  ({\it b}) and spin-spin correlation 
parameter $C_{y,y}$ ({\it c}) of the  reaction $p+d\to (pp)_s+n$ 
 at the neutron scattering angle $\theta_{c.m.}^n=180^0$.
  Curves show the results of calculations
 with the RSC potential  at $E_{pp}= 3 $ MeV for the different mechanisms:
ONE without rescatterings  (dashed), ONE  with rescatterings (dotted),
 coherent sum ONE+$\Delta $+SS of the ONE with rescatterings, 
$\Delta$ and SS  (full).
In Fig. ({\it a}) the SS- and $\Delta$-contributions are shown
by the dashed-dotted curves.
}
\label{srescat}
\end{figure}

\eject
\begin{figure}
\mbox{\epsfig{figure=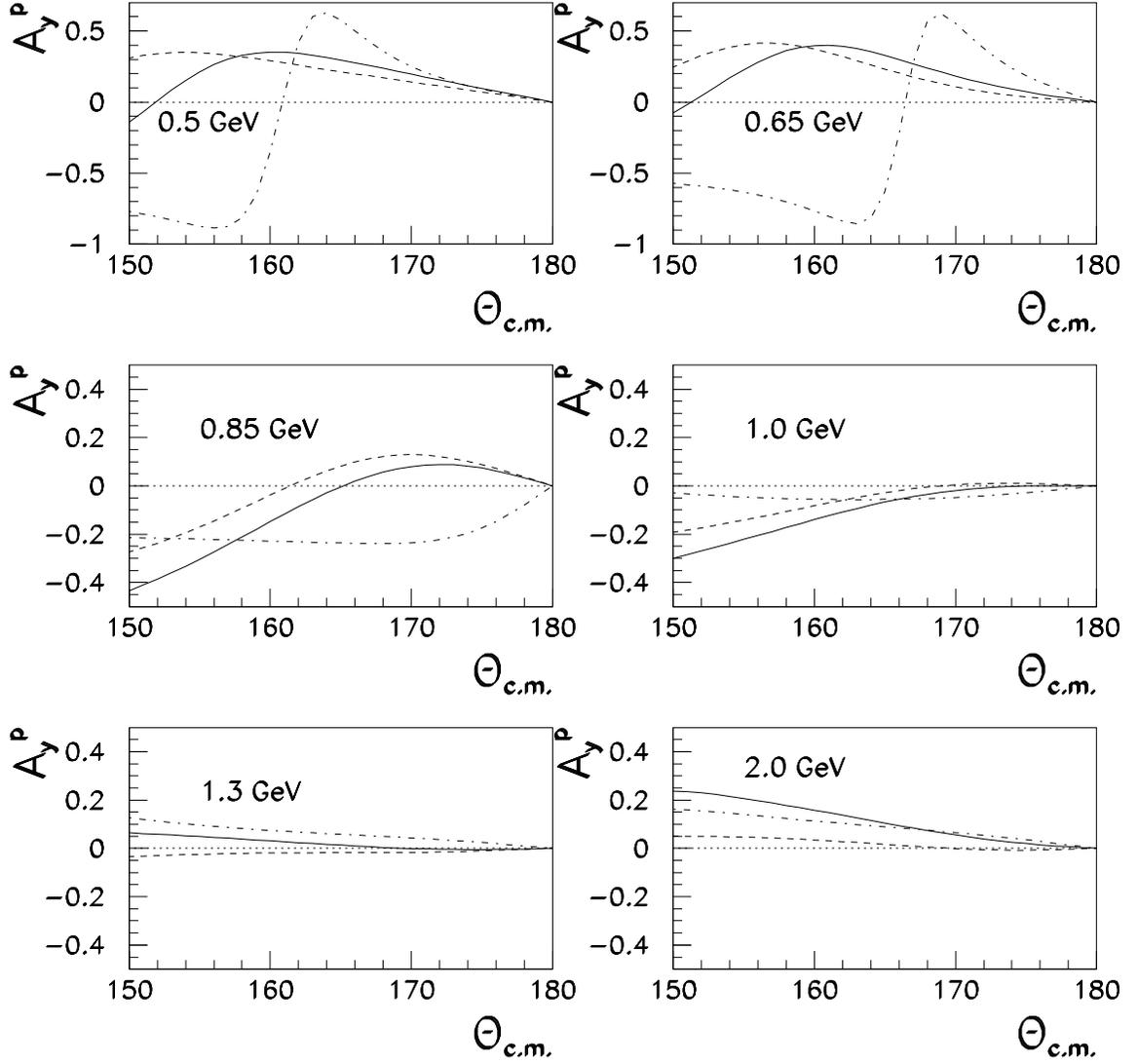,height=0.75\textheight, clip=}}
\caption{ Vector analyzing power of the proton  $A^p_y$ in the
 ${\vec p}+ d\to (pp)_s+n$ reaction at $E_{pp}=3$ MeV
 for the different mechanisms at kinetic energies
 $T_p=0.5$, 0.6, 0.85, 1 , 1.3 and 2 GeV versus 
the c.m.s. scattering angle of the neuteron: one-nucleon
 exchange with rescatterings,
ONE(DWBA), (dashed-dotted); $\Delta$ ( dotted);
 ONE+$\Delta$+SS  (dashed) ;
 ONE(DWBA)+$\Delta$+SS ( full);
 }
\label{asym}
\end{figure}

  Numerical calculation for the matrix elements in the numerators
 and denominators of the ratios (\ref{xratioone}),
 (\ref{xratiod})
 are performed for
 the RSC potential of the NN-interaction. Some results are shown in
 Fig.\ref{xratio}.  One can see
  that there is a small enhancement factor for the singlet pair
  caused by the difference of the spatial parts of the
 $(np)_s/(np)_t$-production  amplitudes.
 This factor is approximately the
 same for the the ONE- (at $T_0>1 $GeV), $\Delta$ and SS-mechanisms,
$ R_X(ONE)\sim R_X(\Delta)\sim  R_X(SS)\sim 1.2-1.5$.
The similar spatial  factor is obtained here
for the $N^*$-excitation mechanism too.
 The minimums  at  $T_0=0.8$ GeV and $T_0\sim 0.75$ GeV
 display the nodes of the deuteron  S-wave function
 $u(q)$ at $q\sim 0.4 GeV/c$
 and the  half-off-shell t-matrix (\ref{half}) at $q'\sim 0.38$GeV/c,
 respectively. The node of the wave function  $u(q)$
 is nonvisible in the cross sections of
 the processes  $pd\to dp$, $p\,d\to (np)_t\,p$ and  (\ref{reaction1})
 due to contribution of the D-wave in the deuteron or $(NN)_t$-scattering
 wave function \cite{smuz98}.

 One should note that the  node of the deuteron S-wave function  $u(q)$ was
 dedicated  indirectly only  by means of
 very complicated experiments for measurements of the tensor analyzing
 power $T_{20}$ in elastic electron-deuteron scattering
 and extraction
 of the deuteron charge formfactor
 (see  Ref.\cite{nikhef} and references therein).
  The similar node of the half-off-shell $t_s(q',k)$-matrix
  at $q'\sim 0.38 GeV/c$
  was not yet observed experimentally.

 One can see from Fig.\ref{srescat}({\it a}) that the ONE mechanism dominates
 in the cross section of the  breakup $p\,d\to (pp)_s\,p$ excluding
 the  region at $T_p\sim 0.75$ GeV.
 Therefore the cross section of the reaction
(\ref{reaction1}) is proportional to $\sim \left [ u^2(q)+w^2(q)\right ]
\times |t_s(q',k)|^2$ at high momenta $q \sim q'$. The  node of
 the half-off-shell t-matrix (\ref{half}) at $q'\sim 0.38$GeV/c
  manifests itself  as a deep  in the cross section
  at $T_0=0.75 $GeV.
 The role of $\Delta $-mechanism is nonnegligible  in this node region only.
   A minor contribution
 is expected for  the mechanism of  excitation  of heavier nucleon
 isobars $N^*$ at $T_0>1 GeV$ due to approximately the same  factors
  $R_X$ and $R_I$
\footnote{An only exclusion might be the $N^*(1535)$-isobar strongly
coupled to the $\eta-$meson which   maximal
 contribution in the reaction  (\ref{reaction1}) is
 expected at $T_0\sim 1.6 $ GeV.}.
 The maximal value of the cross section of the reaction (\ref{reaction1}) is
 expected at $E_{pp}=0.3-0.7$ MeV.
 At $E_{pp}<0.3 MeV$ the cross section becomes smaller due to
 Coulomb repulsion in the pp-system.

 As shown in Refs. \cite{lak,imuzsh89}, \cite{uz98}, the SS mechanism
 is not important in the  $pd\to dp$ cross section. The relative
 role of the
 SS-mechanism  increases in the reaction (\ref{reaction1}). Nevertheless
 the  SS and $ \Delta$-mechanisms do not fill in the remarkable
 deep minimum of the cross section caused by the ONE mechanism
 (Fig.\ref{srescat}({\it a})). This minimum can be smeared only at higher
 relative energies $E_{pn}> 5-10$ MeV due to contribution of the states
 with nonzero orbital momenta $l\not=0$ in the half-off-shell NN-amplitude
 \cite{smuz98}.
 There is a full dominance of the ONE mechanism
 in the  observables
at $T_0>1 $GeV.

 Rescatterings in the initial and final states fill in in part
 the ONE-minimum of the cross section.
  Nevertheless,  this minimum is visible enough
(Fig.\ref{srescat}({\it a})).
  As  shown in Fig.\ref{srescat}({\it b,c}),
  rescatterings produce a remarkable structure both in $T^{ONE}_{20}$
 and in the spin-correlation parameter $C_{yy}$ (for the definition see
 Ref.\cite{ohlssen})
  in the region of node of the half-off-shell NN-amplitude.
  We stress that in the nodeless case of the $pd\to dp$
  process the similar rescatterings do not change practically
  the tensor analyzing power $T^{ONE}_{20}(180^o)$
  \cite{uz97}.
 Therefore  rescatterings give an additional signal for the presence
 of the node.

The $A_y^p$ value
is determined by interference of three  contributions ONE+$\Delta$+SS
at energies $T_p\leq$  1GeV and predominantly by the ONE at $T_p > 1$GeV.
Each of this mechanisms alone results in zero value of the $A_y^p$ , but
as it seen from Fig. \ref{asym}, their interference provides the 
nonvanishing values of the $A_y^p$ at polar angles of the pair ejection 
even close to $\theta_{pp}\sim 0^o$.
The one-nucleon exchange  mechanism with rescatterings denoted as ONE(DWBA)
in Fig.\ref{asym},
  gives nonzero values $A_y^p$ also.

 The reaction (\ref{reaction1}) was not yet investigated experimentally.
 There is available a rather poor experimental information  about the
 squared amplitude for singlet
 pair formation $|A(p\,d\to (pn)_s\,p)|^2$ obtained in semiinclusive
 experiments only
 \cite {dubna,bfw, boudard}.
  The  energy resolution  in Refs. \cite{dubna, boudard} was not
 enough high to observe on the basis of relation Eq.(\ref{contin})
 the important properties of the breakup
 $p\,d\to (pn)_s\,p$ caused by the ONE-mechanism.

 In conclusion, we have found that the broad class of the diagrams
 related to excitation of the nucleon isobars,  which
 are important in the process $pd\to dp$, are diminished
 by the isotopic spin relations in the singlet channel
 of the deuteron breakup.
 As a result the relative contribution of the  ONE mechanism increases
  essentially
 in  the reactions (\ref{reaction1}) and   $p\,d\to (pn)_s\,p$. Moreover,
 the ONE mechanism  becomes more interesting in this channel due to
 suppression of the  $l \not= 0$-states in the half-off-shell
 NN-amplitude at low $E_{NN}$.
   Rescatterings in the initial and final states
 do not smear the remarkable minimum of the cross section  caused
 by the node of the $t(q,k)$-amplitude and display its presence
 in a specific behaviour of $T_{20}$ and $C_{yy}$.
 It allows to obtain from these reactions at $T_0=0.5-3 GeV$
 an  information about the deuteron wave function $u(q)$
 and half-off-shell amplitude of the NN-scattering $t_s(q',k)$ in the
 $^1S_0$-state  at high internal momenta  $q,q'\sim 0.3 - 0.65$GeV/c.

\eject
 {\bf Acknowledgments}. Author is thankful to
 Prof. H. Str\"oher and Dr. H. Seyfarth
 for reading manuscript and many fruitful discussions
 and also to Profs. V.I. Komarov,  L.A. Kondratyuk and other members
 of ANKE-collaboration for interesting   remarks.
 Author is thankful to the direction of the IKP-II KFA
 for warm
 hospitality during the visits to J\"ulich were  a part of this work
 was done. This work was supported  by the WTZ grant
 KAS-001-099.


\end{document}